\documentclass[fleqn,usenatbib]{mnras}

\usepackage{newtxtext,newtxmath}
\usepackage[T1]{fontenc}

\DeclareRobustCommand{\VAN}[3]{#2}
\let\VANthebibliography\thebibliography
\def\thebibliography{\DeclareRobustCommand{\VAN}[3]{##3}\VANthebibliography}

\usepackage{graphicx}	% Including figure files
\usepackage{amsmath}	% Advanced maths commands
\usepackage{bm}
\usepackage[normalem]{ulem}

\title[Plasma Drift in Pulsars]{Polar Cap Region and Plasma Drift in Pulsars} 

\author[Szary et al.]{
Andrzej Szary$^{1}$\thanks{E-mail: a.szary@ia.uz.zgora.pl}
and  Joeri van Leeuwen$^{2}$
\\
$^{1}$Janusz Gil Institute of Astronomy, University of Zielona G\'ora, Licealna 9, 65-417 Zielona G\'ora, Poland\\
$^{2}$ASTRON, the Netherlands Institute for Radio Astronomy, Postbus 2, 7990 AA, Dwingeloo, The Netherlands
}

\date{Accepted XXX. Received YYY; in original form ZZZ}

\pubyear{2023}

\begin{document}
\label{firstpage}
\pagerange{\pageref{firstpage}--\pageref{lastpage}}
\maketitle

\begin{abstract}
Pulsars often display systematic variations in the position and/or intensity of the subpulses, the components that comprise each single pulse. 
Although the drift of these subpulses was observed in the early years of pulsar research, and their potential for understanding the elusive emission mechanism was quickly recognised, there is still no consensus on the cause of the drift.
We explore the electrodynamics of two recently proposed or refined drift models: one where plasma lags behind corotation, connecting the drift with the rotational pole; and another where plasma drifts around the electric potential extremum of the polar cap.
Generally, these are different locations, resulting in different drift behaviours, that can be tested with observations.
In this study, however, we specifically examine these models in the 
axisymmetric case, where the physics is well understood.
This approach seems counter-intuitive as both models then predict similar large-scale plasma drift. 
However, it allows us to show, by studying conditions \emph{within} the sparks for both models, that the lagging behind corotation (LBC) model is inconsistent with Faraday's law.
The modified carousel (MC) model, where plasma drifts around the electric potential extremum, not only aligns with Faraday's law, but also provides a future direction for developing a comprehensive model of plasma generation in the polar cap region.
Unlike previous models, which considered the drift only inside the discharging regions, the MC model reveals that the electric field \emph{between} the discharges is not completely screened, and plasma drifts there -- a paradigm shift for the drifting subpulse phenomenon.
%\vspace{1cm}

%This is a simple template for authors to write new MNRAS papers.
%The abstract should briefly describe the aims, methods, and main results of the paper.
%It should be a single paragraph not more than 250 words (200 words for Letters).
%No references should appear in the abstract.
\end{abstract}


%%%%%%%%%%%%%%%%%%%%%%%%%%%%%%%%%%%%%%%%%%%%%%%%%%

%%%%%%%%%%%%%%%%% BODY OF PAPER %%%%%%%%%%%%%%%%%%
\section{Introduction}

The drifting-subpulse phenomenon, a feature of pulsar emission where the single-pulse components, the so-called subpulses, exhibit systematic variation in position or intensity, or both, is still an unsolved problem.
Although this striking behaviour was discovered in the early years of pulsar research  \citep{1968_Drake},  there is still no consensus regarding a main feature of this drift: the  direction of plasma motion in the pulsar magnetosphere.

In the force-free state, with the Lorentz force zero, the magnetosphere is filled with charged particles and the electric field $\bf E_{\perp}$ perpendicular to the magnetic field $\bf B$ is
\begin{equation}
    {\bf E_{ \perp} } = -({\bf \Omega} \times {\bf r}) \times {\bf B} / c, 
    \label{eq:ff}
\end{equation}
where $\bf \Omega$ is the angular velocity, $\bf r$ is the location vector, and $c$ is the speed of light.
In this idealised force-free state the particles would corotate with the star with velocity \mbox{${\bf v_{\rm cor}} = {c\left ( \bf {E}_{\perp} \times B \right)}/{ B^2}$}.
In the acceleration region, by definition, the density of charges is less than the corotational charge density.
Plasma in this region is both accelerated along the magnetic field and moves perpendicular to the magnetic field with velocity:
\begin{equation}
 {\bf v} = \frac{c(\bf \widetilde{E}_{\perp} \times B)}{B^2},
\end{equation}
where ${\bf  \widetilde{E}_{\perp}} $ is the electric field in the plasma starved region perpendicular to the magnetic field.
The direction and magnitude of ${\bf  \widetilde{E}_{\perp}} $ determine the plasma drift direction in the pulsar magnetosphere.
Models of plasma generation assume that the pair cascade is localised in the form of discharges (called \textit{sparks}) over the polar cap which produce plasma columns along the open magnetic field lines.
The observed radio emission is a consequence of this non-stationary plasma flow, and this emission is formed at altitudes of ${\sim}500$~km above the neutron star surface \citep[see, e.g.,][]{2000_Melikidze, 2003_Kijak, 2017_Mitra}.

% pdftoppm fields.pdf fields -png -r 300
\begin{figure*}
    \centering
    \includegraphics[width=470pt]{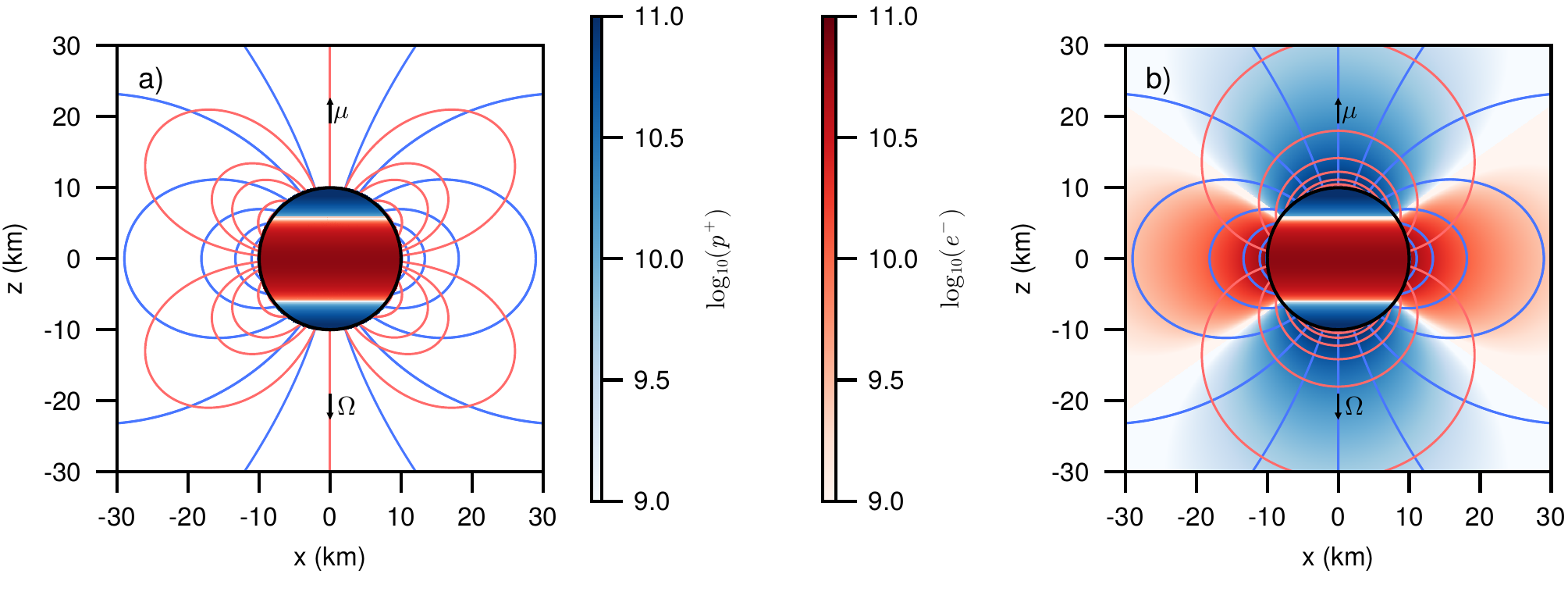}
    \caption{Vacuum ({\it panel (a)}) and force-free ({\it panel (b)}) electromagnetic field of an anti-align dipole rotator. 
    Blue lines show the magnetic field lines, while red lines correspond to the electric field lines. 
    The colorbars show the number charge density at the neutron star surface or at plane $y=0$ beyond the neutron star.}
    \label{fig:fields}
\end{figure*}

In their pioneering paper, \cite{1975_Ruderman} propose that the above mentioned perpendicular plasma drift forces the sparks to rotate in a `carousel` around the magnetic axis. This explains drifting subpulses, and a variety of further pulsar data.
In their first estimate of the drift speed, \cite{1975_Ruderman} assume  that a) the accelerating potential peaks at the magnetic axis, and that b) it falls off linearly to zero at the edge of the polar cap. While these are very valid starting points, they have since been refined to be more physical. A more general model that does not require simplification b) was presented in \citet{2012_Leeuwen}, and in \citet{2017_Szary} we also improve on a), showing that sparks rotate not around the magnetic axis \textit{per se}, but around the location of the electric potential extremum of the polar cap.
If this potential extremum coincides with the centre of the polar cap, the subpulses seem to drift around the magnetic axis -- even for a non-dipolar configuration of the surface magnetic field. 
This  Modified Carousel (hereafter, MC) model can be used to interpret the most extraordinary drifting behaviours like, for instance, the bi-drifting phenomenon \citep[][]{2017_Szary, 2020_Szary} or the drift direction changes in PSR J1750$-$3503 \citep{2022_Szary}.

In  recent years, an alternative model, where drift is produced as plasma lags behind the corotation of the neutron star, gained momentum \citep[see, e.g.][]{2016_Basu, 2020_Mitra, 2020_Basu}.
This Lagging Behind Corotation (hereafter, LBC) model connects the drift with the rotation axis, in clear contrast to the carousel models.
Despite very different assumptions about the plasma drift direction, the LBC model is equally capable of describing some extraordinary drifting behaviour, for instance, the bi-drifting phenomenon \citep{2020_Basu, 2022_Basu}.

In general there are two sources of electric field in a rotating magnetosphere: a potential electric field due to the charge density in the magnetosphere, and an inductive field due to the time-changing magnetic field.
Since the electrodynamics of an oblique rotator is still an unsolved problem \citep{2016_Melrose}, we test the drift models for the case of an aligned rotator, for which the physics is well understood.

The paper is organised as follows.
In Section \ref{sec:theoretical} we present theoretical background and introduce two models of drifting subpulses studied in this publication.
In Section \ref{sec:electrodynamics} we explore the electrodynamics of the drifting models.
The discussion is in Section \ref{sec:discussion}, followed by conclusions presented in Section \ref{sec:conclusions}.

\begin{figure*}
    \centering
    \includegraphics[width=490pt]{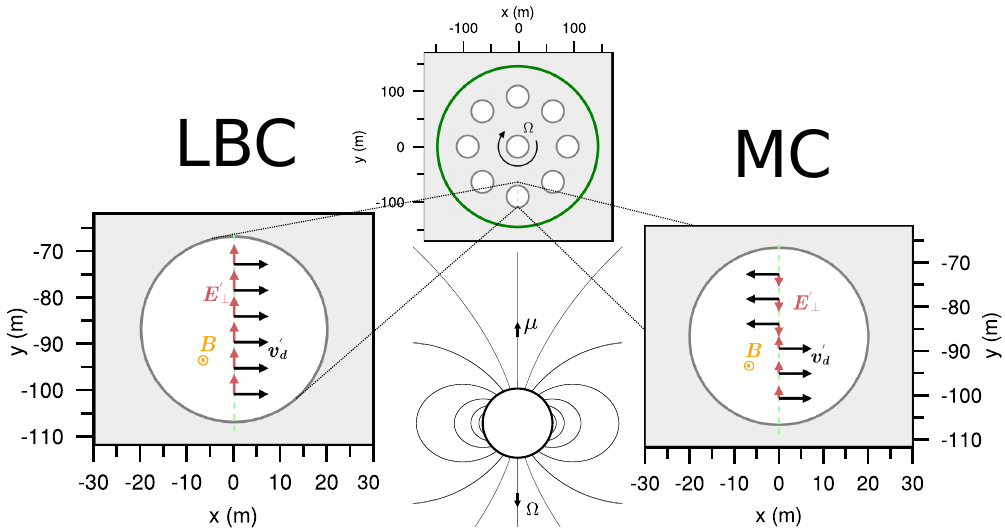}
    \caption{Top-down view of the configuration of sparks at the polar cap ({\it the middle panel}). 
    The polar cap boundary is shown with a green circle, while the white circles correspond to the spark-forming regions. 
    {\it The left and right panels} show a single spark-forming region for LBC and MC models, respectively. 
    The orange arrow {$\odot$}, coming out of the Figure, represents the magnetic field while  the red arrows indicate a component of electric field perpendicular to the magnetic field. 
    The black arrows represent the drift velocity. 
    All the quantities are show in the corotating frame of reference. 
    The grey regions correspond to the regions where plasma corotates with the star.}
    \label{fig:polar_cap}
\end{figure*}

\section{Theoretical background}
\label{sec:theoretical}

A pulsar can be described by its magnetic moment $\bm \mu$ and angular velocity $\bm \Omega$.
In the paper at hand we consider the ''anti-aligned rotator'' where ${\bm \mu}$ is parallel to ${\bf \Omega}$ but with opposite directions.
This relatively simple case has a few characteristics one needs to be aware of.
First, in this geometry the rotational and magnetic poles are coincident, which means we cannot distinguish between them
in practice. If the extremum from the MC model is located at the magnetic pole, it will also be equal to the rotation
centre of the LBC model.
Second, in reality such a ``pulsar'' would not straight-forwardly pulsate, as the configuration is symmetric about the
rotation axis\footnote{Although, as a secondary effect, subpulses could still be present, and a periodic modulation of
those components could be detectable.}.
That said, for our study, below, of the electrodynamics of the drifting models, the aligned rotator is the best place to start
because of its relatively fundamental behavior.

In the early years of pulsar astrophysics, a star surrounded by vacuum was considered \citep[see, e.g., ][]{1967_Pacini, 1968_Pacini, 1969_Ostriker}.
Since the neutron star is an almost perfect conductor, the electric field inside the star as measured in the corotating frame of reference becomes negligible.
It determines the non-zero internal electric field in the observer's frame of reference, which is a result of charge re-distribution to balance out the Lorentz force.
For the anti-aligned rotator this gives rise to an excess of negative charges at the equator and positive charges at the poles.
In general the net charge of the star is the sum of the interior and surface charges.
The vacuum electromagnetic field of an anti-aligned rotator with zero net charge in spherical coordinates ($r$, $\theta$, $\phi$) is given by \citep{1955_Deutsch, 2017_Cerutti}
\begin{equation}
    \left ( B_r, B_\theta, B_\phi \right ) = B_{\rm eq}\left ( \frac{R}{r} \right )^3 (2\cos{\theta}, \sin{\theta}, 0),
\end{equation}
\begin{equation}
    \left ( E_r, E_\theta, E_\phi \right ) = \frac{\Omega R}{c} B_{\rm eq}\left ( \frac{R}{r} \right )^4 (1 - 3\cos^2{\theta}, -\sin{2\theta}, 0),
    \label{eq:efield_vac}
\end{equation}
where $R$ is the stellar radius, and $B_{\rm eq}=\mu / R^3$ is the magnetic field strength at the equator.
In Fig.~\ref{fig:fields}a) we show the magnetic and electric fields for an anti-aligned rotator with  net charge equal to zero, surrounded by vacuum.

While the vacuum model assumes that the strong neutron-star gravity traps free particles at the stellar surface,  Equation \ref{eq:efield_vac} shows there is a strong unscreened parallel component of the electric field $E_{\parallel}$.
\cite{1969_Goldreich} have shown that this electric force greatly exceeds the gravitational force and leads to lifting of particles from the surface and pair creation in the magnetosphere.
The charge density necessary to cancel $E_{\parallel}$ is called the Goldreich-Julian (GJ) density, which  near the neutron star is given by
\begin{equation}
    \rho_{\rm GJ} \approx -\frac{\bf \Omega \cdot B}{2 \pi c} = \frac{B_{\rm eq}}{2 \pi c} \left ( \frac{R}{r} \right ) ^3(3\cos^2{\theta} - 1).
\end{equation}
With this charge density, the electric field in the neutron star and in its magnetosphere vanish in the corotating frame of reference, while in the observer's frame it is described by Equation \ref{eq:ff}.
Figure \ref{fig:fields}b) shows the electromagnetic field and charge density for a force-free magnetosphere.
Such a force-free magnetosphere corotates with the star in the region where the closed field lines extend to the light cylinder.
The light cylinder is defined as the cylindrical radius $R_{\rm LC} = c/\Omega = cP/(2\pi)$ at which the corotation speed equals the speed of light.
The so-called open magnetic field lines occupy the region above the polar caps, with the last open field line crossing the stellar surface at the polar angle $\theta_{\rm pc}$ given by $\sin^2{\theta_{\rm pc}=R / R_{\rm LC}}$.
Early particle-in-cell (PIC) simulations of pulsar magnetospheres suggest that the particle acceleration, and thus electron-positron creation, is suppressed in most parts of the polar cap \citep[see, e.g.,][]{2014_Chen, 2015_Philippov_a}.
However, including general relativistic effects, especially the frame-dragging of space-time by the stellar rotation, ignites the $e^{\pm}$ discharge above the polar cap for a sufficiently high compactness of the neutron star \citep{2015_Philippov_b, 2016_Gralla, 2016_Belyaev}.
The cascade pair production happens just above the polar cap, in a quasi-periodic manner \citep{2013_Timokhin}.

\subsection{Two models of drifting subpulses}

A model explaining drifting phenomenon was proposed early on in \cite{1975_Ruderman},  describing how sparks drift around the magnetic axis.
This seminal model does, however, have room for further improvement and correction. This is especially the case for two aspects.
First, the center of the spark rotation is assumed to be the magnetic axis. 
Although this a reasonable first-order approximation, there is no intrinsic reason why the carousel should spin around this axis. This is especially problematic if we consider non-dipolar surface magnetic fields, as favoured by recent observations of X-ray emission from the stellar surface \citep[see, e.g.,][]{2008_Gil, 2017_Szary_b, 2018_Hermsen, 2020_Petri}. 
Secondly, this assumption means that for an inclined rotator, parts of the polar cap contain sparks that move faster than the corotating magnetosphere.
Both of these concerns were alleviated in the MC model \citep{2017_Szary}. 
We have shown that if the accelerating electric field ($E_\parallel$) in regions between sparks is not fully screened, plasma in those regions drifts around the electric potential extremum of the polar cap \citep[see, e.g., Fig.~8 in][]{2017_Szary} -- not around the magnetic axis, per se.
If the potential extremum coincides with the centre of the polar cap, and radio emission is generated at higher altitudes where the magnetic field has a dipolar configuration, subpulses appear to drift around the magnetic axis even for a case of non-dipolar surface magnetic field.
Furthermore, we have shown that, for instance, inside the spark-forming region the generated plasma circulates around the spark centre which means, that in the plasma starved regions the generated particles can move faster than the corotating magnetosphere \citep[see Fig.~6 in][]{2017_Szary}.

\begin{figure*}
    \centering
    \includegraphics[width=460pt]{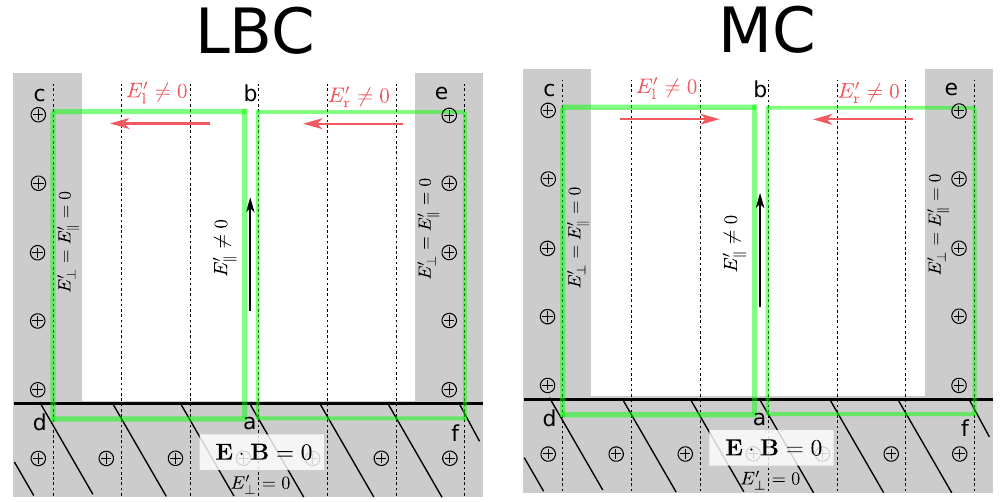}
    \caption{Side view of a single spark-forming region for the spark configuration presented in Fig.~\ref{fig:polar_cap}.
    The grey zones corresponds to regions where plasma corotates with the star.
    {\it The left panel} corresponds to the LBC model, while {\it the right panel} corresponds to the MC model.
    The parallel and perpendicular components of electric field are shown with black and red arrows, respectively.}
    \label{fig:circulation}
\end{figure*}

The LBC model, on the other hand, describes  the sparks as lagging behind the rotation of the pulsar, thus connecting the drift with the rotation axis \citep{2020_Mitra, 2020_Basu}.
The model was further developed using insights from the partially screened gap model \citep{2003_Gil}. 
In this model, the gap is not a complete vacuum: it is partially screened by a steady supply of positively charged ions, extracted from the surface.
This scenario requires the continuous presence of sparks bordering the boundary of the polar cap.
The sparks are as tightly packed as possible which leads to two distinct directions of spark movement across the polar cap: one half of the sparks shows a clockwise shift while the other half moves anticlockwise  \citep[see Fig.~3 in][]{2022_Basu}.

\section{Electrodynamics of the {polar cap region}}
\label{sec:electrodynamics}

\subsection{Corotating frame of reference}

To study plasma drift in pulsars we use the corotating frame of reference, where plasma corotating with the star is at rest. 
In order to describe the electrodynamics in this frame the methods of general relativity should by used.
Similar to mechanical analyses where rotating frames include ''fictitious'' forces, the electrodynamics analyses include ''fictitious'' charges and currents.
The Gauss law in the rotating frame is \citep{2016_McDonald}

\begin{equation}
    \nabla^{\prime} \cdot \mathbf{E}^{\prime} = 4 \pi \rho^{\prime} + \frac{2 \mathbf{\Omega} \cdot \mathbf{B}^{\prime }}{c} - \frac{\mathbf{v}}{c} \cdot \nabla^{\prime} \times \mathbf{B}^{\prime}, 
	\label{eq:gauss}
\end{equation}
while Faraday's law in the rotating frame is given by \citep{2016_McDonald}
\begin{equation}
    \nabla^{\prime} \times \mathbf{E}^{\prime} = - \frac{1}{c} \frac{\partial \mathbf{B}^{\prime}}{\partial t^{\prime}}.   
	\label{eq:faraday}
\end{equation}
Here $\rho^{\prime}$ is the charge density in the rotating frame and $\bf v = \Omega  \times r$ is the velocity in the observer's frame. 

\subsection{Electric field in polar cap region}
\label{sec:electric_field}

The MC and LBC models assume a different direction for the electric field in the plasma starved regions, resulting in a different plasma drift direction.
In Fig.~\ref{fig:polar_cap} we show this difference, for an example polar cap configuration for a pulsar with $P=1 \,{\rm s}$, $\dot P = 10^{-15} \, {\rm s/s}$ and a dipolar surface magnetic field  configuration.
The polar cap, visible in the middle panel, is populated with nine sparks with radius 20 meters (one at the centre and eight on a circle of radius 90 meters).
Because the rotational pole and the potential extremum are co-located in this example, the large-scale drift behaviour of the sparks across the polar cap in the middle plot is the same for both models.
The side panels show the top-view of a single spark-forming region, for both the LBC and MC models, and this is where the difference arises that we explore in this work.
All the quantities in the Figure are show in the corotating frame of reference, 
and we here assume the electric field is entirely screened in the areas between the sparks.

In Fig.~\ref{fig:circulation} we show a side view of a single spark-forming region for the LBC (the left panel) and MC (the right panel) models.
Since for an anti-aligned rotator $\partial \mathbf{B}^{\prime} / \partial t^{\prime} = 0$ with high accuracy \citep[see, e.g.,][]{2012_Leeuwen}, the circulation of the electric field along a closed path should be zero.
Let us consider two paths $a b c d a$ and $a b e f a$. 
The lateral sides ($c d$ and $e f$) follow magnetic field lines in the regions beyond a spark, while the top parts ($b c$ and $b e$) cross the acceleration gap; the bottom parts ($d a$ and $f a$) are just below the neutron star surface.
Here we consider an idealised case where the neutron star is a perfect conductor ($E^{\prime}_\perp=0$) and there is no acceleration along magnetic field lines outside of a spark ($E^{\prime}_\parallel=0$).
The circulation of the electric field along the paths is

\begin{equation}
\begin{split}
    \oint_{abcda} {\bf E^{\prime} \cdot dl^{\prime}}  & = \int_a^b {\bf E_\parallel^{\prime} \cdot dl^{\prime}} + \int_b^c {\bf E_l^{\prime} \cdot dl^{\prime}}  = 0 \\ 
    \oint_{abefa} {\bf E^{\prime} \cdot dl^{\prime}}  & =  \int_a^b {\bf E_\parallel^{\prime} \cdot dl^{\prime}} + \int_b^e {\bf E_r^{\prime} \cdot dl^{\prime}} = 0, 
\end{split}
\end{equation}
where $E_\parallel^\prime$ is the parallel component of electric field, while $E_l^{\prime}$ and $E_r^{\prime}$ are perpendicular components for the left and right sides of the spark-forming region under consideration, respectively.
For the considered paths the first term is always positive while the second term is always negative except the $b c$ part in the case of the LBC model.
This shows that the direction of the perpendicular component of the electric field predicted by the LBC model is inconsistent with Faraday’s law.

\begin{figure*}
    \centering
    \includegraphics[width=465pt]{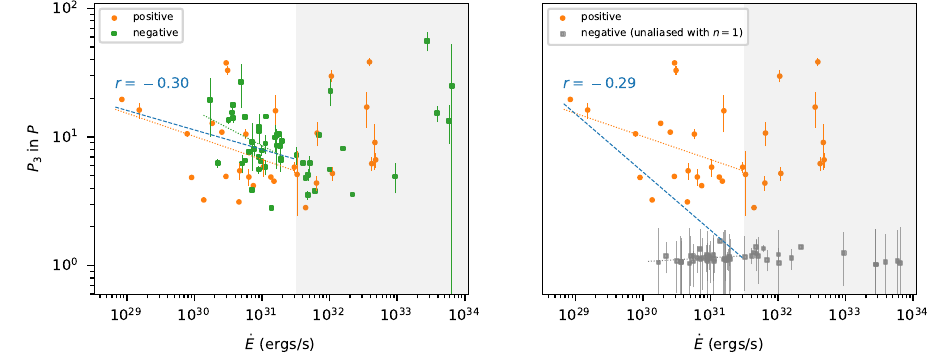}
    \caption{Drifting periodicity $P_3$ as a function of the rotational energy-loss rate  $\dot{E}$ for pulsars from the \citet{2023_Song} MeerKAT sample with clear phase variations. In orange, circles correspond to positive drift and the dotted line is their best-fit powerlaw. In the left-hand panel, the green squares and dotted line correspond to negative drift; in the right-hand panel, these  negative drifters are corrected for an hypothesised alias order $n=1$, and shown as gray squares with a gray dotted line. In blue, we also show the correlation coefficient $r$ and best-fit powerlaw (dashed line) of the region log$(\dot{E})<31.5$.  }
    \label{fig:p3edot}
\end{figure*}

In a more realistic case, the electric field in the areas between sparks may not be fully screened.
In fact, for the MC model it is required in order to enforce plasma in those regions to drift around electric potential extremum at the polar cap.
However, even when we consider such scenario, the LBC model is still inconsistent with Faraday's law as it would require accelerating electric field in regions between sparks to be greater than inside spark forming regions.

\section{Discussion}
\label{sec:discussion}

The phenomenon of subpulse drifting is present in $\sim 60 \%$ of the overall pulsar population \citep{2023_Song}, and below we discuss the implications of our modelling findings for the observed sample.
Initially, the main  framework for interpreting the observed subpulse drift is the carousel model as  proposed by \cite{1975_Ruderman}.
As mentioned before, that model assumes for simplicity that sparks drift around the magnetic axis. 
%While not unreasonable, there is little solid physical justification that this axis is indeed the centre of the subpulse motion for all pulsars. 
At that time, a few tens of pulsars with subpulse drift were known \citep{1975_Ritchings}.
With an advent of new high-quality pulsar data the number of observed drifters has increased significantly since, well into the hundreds, with much higher phase-resolved resolution, and significantly improved detail on the time variability.
This development sparked new interest in untangling this  unsolved problem in pulsar astronomy.
In recent years progress in this field has gone in two directions, which we discuss below.

%%%%%%%%%%%%%%%%%%%%%%%%%%%%%%%%%%%%%%%%%%%%%%%%%%%%%%%%%%%%%%%%%%%%%%%%%%
% LBC
In \cite{2016_Basu} the LBC model, which connects the drift with the rotation axis, was introduced.
Two driving factors were at play.
Since charges in the pulsar magnetosphere are responsible for generating the electric field responsible for corotation (see Equation \ref{eq:ff}) it is reasonable to assume that in charge-starved regions the plasma should lag behind the corotation.
This lag translates to subpulses always intrinsically drifting from the trailing to the leading edge of the profile. The \emph{observed} drift direction next depends on the aliasing. 
Second, the authors found an anticorrelation of the repetition time of the drift pattern ($P_3$) with the rotational energy loss ($\dot{E}$), 
for the low $\dot{E}$ regime. 
For the studied sample of 27 pulsars with phase-modulated drifting, 
the anticorrelation seems to be more evident when assuming the negative drifters are aliased, 
as hypothesized by the LBC model.
Here, the positive drift -- from earlier to later longitudes -- is interpreted as an unaliased measurement, while the negative drift -- from later to earlier longitudes -- is interpreted as an aliased measurement. 
\citet{2016_Basu} state the $P_3-\dot{E}$ correlation thus found  corroborates the LBC model.

The sample of phase-modulated drifting pulsars has recently been significantly expanded 
using MeerKAT \citep{2023_Song}, allowing us to revisit the claimed correlation with improved statistics.
In Fig.~\ref{fig:p3edot} 
we show the relevant sample of $P_3$ measurements published in \cite{2023_Song}. 
The sample is made up by the 81 pulsars with clear drift direction: 31 pulsars with positive and 50 pulsars with negative drift.
In the left panel, these are shown \emph{as is}, which the MC model allows.
In the right panel we assume all negative drifters are aliased, as in the LBC model they must.
There, the grey squares represent the unaliased value of $1 / P_3= n + 1 / P_3^{\rm obs}$ for negative drifters, assuming the alias order $n=1$. 
Here $P_3^{\rm obs}$ is the apparent (possibly aliased) modulation in units of pulse periods.
This panel thus represents the true drift required under the LBC model.
We compare how well the drift rates correlate with $\dot{E}$ by calculating the Pearson correlation coefficient $r$. 
We determine the strength of the correlation for the region \mbox{log$(\dot{E})<31.5$}, 
 where \citet{2023_Song} found $P_3$ decreases with increasing $\dot{E}$, before turning up. 
In both cases we find weak, negative correlations of $r$=$-$0.3.
As these are of equal strength, we
 conclude there is no evidence to favour the LBC interpretation that the negative drifters represent aliased measurements,
over accepting the drift values as observed.
On the contrary, for the observed values (left panel of Figure~\ref{fig:p3edot}), the slopes of powerlaws  separately fitted to the negative and positive subsets are consistent with one another (dotted lines). However, the powerlaw coefficient for the positive drift is $-0.18$ while for the unaliased values for negative drift it is $0.03$ (grey dotted line in right panel of Figure ~\ref{fig:p3edot}).
Under the LBC model, these values should also agree. 

In their Appendix, \cite{2020_Mitra} used the Gauss law to show that the drift direction is opposite to the corotation velocity.
The authors argue that using the assumption that $E_r^\prime$ does not change significantly across the gap results in $\iint_S E^{\prime}_r dS = 0$.
However, $E_r^\prime$ changes significantly along the gap \citep[see, e.g.][]{2010_Timokhin} which results in $\iint_S E^{\prime}_r dS \neq 0$.
Moreover, it was argued that $E_{\theta}^{\prime}$ is directed in the direction behind the linear velocity ${\bf v_d}^{\prime}$, while for the considered case \mbox{$E^\prime_\phi=0$},  $E_{\theta}^{\prime}$ should be perpendicular to ${\bf v_d}^{\prime}$.
More details of electric field in the LBC model can be find in \cite{2020_Basu}, where it is argued that $E_\perp^\prime=-E_\perp$.
It is worth noting that this means that in the observer frame $E_\perp = 0$ everywhere within the spark.
As shown in Section \ref{sec:electric_field} that is inconsistent with Faraday's law.
Note that not only the value of this electric field is problematic, but the fact that in the corotating frame of reference it does not change its direction across the spark.

The LBC model was further developed using insights from the partially screened gap model, resulting in a model of two-dimensional evolution of spark discharges at the polar cap \citep{2022_Basu}.
The model uses the constraints from the polar cap boundary and the assumption that sparks during their lifetime lag behind corotation, which results in the differential shift in the spark pattern in the two halves of the polar cap.
The tightly packed nature of the sparks implies that the heating location at the centre is stationary, which accounts for the absence of drifting in the central core region of the profile window \citep{1986_Rankin, 2019_Basu}. 
The LBC model along with the non-dipolar nature of the surface magnetic fields was used with success to simulate different, even the most extraordinary, drifting behaviours \citep[see, e.g.,][]{2020_Basu, 2023_Basu}.

%%%%%%%%%%%%%%%%%%%%%%%%%%%%%%%%%%%%%%%%%%%%%%%%%%%%%%%%%%%%%%%%%%%%%%%%%%
% MC
The MC model was introduced as an alternative in \cite{2017_Szary}.
The standard  carousel model \citep{1975_Ruderman} considers a group of localised discharges, in the form of sparks.
It assumes that when a discharge begins at some point on the polar cap, it inhibits the formation of another simultaneous discharges within a distance comparable to the gap height.
This model, however, does not explain how a new spark remembers the location where the previous spark was formed; nor does it present the mechanism that prevents discharging in regions between sparks; while 
finally, and foremost, it does not establish  why the spark  rotation should be around the magnetic axis. 
The MC model was the result of a search for a physical justification for spark rotation around the magnetic axis.
In \cite{2017_Szary} we pointed out that the electric potential growing during the discharge influences also the electric field beyond the spark.
The electric field between the sparks can be screened either due to charge separation, by particle generation via discharges or by inflow of charged particles from the surface or outer magnetosphere.
When this electric field is fully screened, plasma \emph{inside} the spark-forming region rotates around the spark centre and no systematic drift is observed.
We have shown, however, that if this electric field is not fully screened plasma in regions \emph{between} sparks rotates around the point of electric potential extremum at the polar cap: minimum in the pulsar case ($\mathbf{\Omega} \cdot \mathbf{B} < 0$) and maximum in the antipulsar case ($\mathbf{\Omega} \cdot \mathbf{B} > 0$).
Furthermore, even for random distribution of sparks, the electric potential extremum is at the centre of the polar cap (see Figure 7 in \citealt{2017_Szary}).
Thus, since the radio emission is produced at much higher altitudes, where the magnetic field  configuration is dipolar, the subpulses should exhibit drift around the magnetic axis.

The proposed model not only explains why subpulses (not the plasma) appear to drift around the magnetic axis, even for the non-dipolar configuration of the surface magnetic field, but we put forward an alternative scenario of plasma drift in the polar cap region.
In our model the plasma between sparks drifts around the electric potential extremum at the polar cap; and the location of the discharges -- the sparks -- is determined by this drifting plasma.
The observed stability of the subpulse structures suggests that the pattern of discharging regions at the polar cap is also stable.
%The question we do not yet have an answer to is what  the source of the plasma is in the areas in between the discharges.
We do not yet have an answer to the question of the source of the plasma in the areas between discharges.
The possible plasma generation or inflow should be continuous, preventing discharges resulting in radio emission in those regions.
One possible scenario is that there exists a reverse plasma flow, induced by a mismatch between the magnetospheric current distribution and the current injected in the spark-forming regions \citep[see, e.g., ][]{2012_Lyubarsky}.

The MC model, with rotation of sparks around the centre of the polar cap, in a natural way explains the absence of drifting in the central core region of the profile window \citep{1986_Rankin, 2019_Basu}.
The MC model can furthermore, like the LBC model, be used to simulate extraordinary drifting behaviour when assuming a non-dipolar structure of the surface magnetic field  \citep[see, e.g.,][]{2017_Szary, 2020_Szary, 2022_Szary}.

In this paper we consider an aligned rotator to test the assumptions of the LBC and MC models. 
In Section \ref{sec:electric_field} we have shown that in the LBC model either the electric field or the boundary condition must be invalid.
If we,  for instance,  consider a different boundary condition in which the underdense region covers the whole polar cap area, then treating the spark as a test particle would result in drift around the poles. 
However, the stable structures of subpulses suggest that the pattern of sparks at the polar cap is also stable.
The discharge (spark) happens when the accelerating electric field is high enough to produce pairs. 
In order to have localised sparks we need regions with higher and lower (or zero) accelerating electric field.
With the underdense region extending to the rim of the polar cap we would lose the mechanism which is responsible for keeping the sparks localised.
Sparks should then happen in random places where some seed charges (or photons) occur and the accelerating electric field is high enough to produce pairs.

In reality most pulsars are inclined rotators and there are two sources of electric field: a potential field, $E_{\rm pot}$, due to the charge density in the magnetosphere and an inductive electric field, $E_{\rm ind}$, due to time-changing magnetic field as pulsar rotates.
The lack of plasma in the spark forming regions affects only the potential component of electric field, $E_{\rm pot}$.
Using the assumption of the LBC model ($E_{\perp} = 0$ in the plasma starved region or $E_{\perp}' = -E_{\perp}$ in the co-rotating frame) for an inclined rotator results in the electric field in the co-rotating frame in one direction across the whole polar cap.
This again would be inconsistent with Faraday's law, similar to the left panel in Figure \ref{fig:circulation}.

\section{Conclusions}
\label{sec:conclusions}

We have used the axisymmetric case with anti-aligned rotation and magnetic axes to test the LBC and MC models of drift at the level of individual sparks. 
We have shown that the LBC model is inconsistent with Faraday's law, while the MC model is consistent.
The predictions of the MC model -- that plasma in between the sparks drifts around the location of the electric potential extremum of the polar cap; that screening of the electric field in between the sparks is incomplete; and that there is continuous plasma supply in regions between the sparks (from the surface, discharges or from the outer magnetosphere) -- can be used to develop a comprehensive model of plasma generation in the polar cap region.
Understanding the plasma drift for the simple configuration considered in this paper plays an important role in the development of pulsar theory.

\section*{Acknowledgements}
We thank the referee for their comments, which significantly improved the quality of the paper.
This work is supported by the National Science Centre, Poland, under grant 2023/51/B/ST9/00291.
JvL acknowledges funding from Vici research programme `ARGO' with project number 639.043.815, and from CORTEX (NWA.1160.18.316), under the research programme NWA-ORC, both financed by the Dutch Research Council (NWO).

\section*{Data availability}
The data underlying this article will be shared on reasonable request to the corresponding author.



\bibliographystyle{mnras}
\bibliography{biblio}

%\begin{thebibliography}{99}
%\end{thebibliography}

% Don't change these lines
\bsp	% typesetting comment
\label{lastpage}
\end{document}